# Biopsym: a Learning Environment for Trans-Rectal Ultrasound Guided Prostate Biopsies

Thomas JANSSOONE[a], Grégoire CHEVREAU[a,b], Lucile VADCARD[c], Pierre MOZER[b] and Jocelyne TROCCAZ[a,1]

[a] *TIMC-IMAG Laboratory, UMR 5525 CNRS UJF, Grenoble France*
[b] *La Pitié Salpétrière Hospital, Paris, France*
[c] *LSE (Laboratory for Educational Science), UPMF, Grenoble, France*

**Abstract.** This paper describes a learning environment for image-guided prostate biopsies in cancer diagnosis; it is based on an ultrasound probe simulator virtually exploring real datasets obtained from patients. The aim is to make the training of young physicians easier and faster with a tool that combines lectures, biopsy simulations and recommended exercises to master this medical gesture. It will particularly help acquiring the three-dimensional representation of the prostate needed for practicing biopsy sequences. The simulator uses a haptic feedback to compute the position of the virtual probe from three-dimensional (3D) ultrasound recorded data. This paper presents the current version of this learning environment.

## Introduction

Prostate cancer is the most widespread cancer for men and the second cause of death after lung cancer in many countries. It can be suspected from a blood analysis with an abnormal rate of Prostate Specific Antigen (PSA) or from a suspicious digital rectal examination of this gland. However, the only way to confirm a prostate cancer is to detect one or several positives samples from the anatomo-pathological analysis of prostate biopsies. The simplest way to access to the prostate for biopsy is through the rectum (see Figure 1 – left and middle).

Biopsy is performed through a trans-rectal way: the doctor introduces an ultrasound (US) probe into the rectum to visualize the prostate and thus directs the needle to the sites which must be sampled. A mechanical guide attached to the US probe directs the needle to the target. The procedure must follow a well defined protocol in order to take sufficient samples to affirm or cancel the presence of cancer. One very standard protocol is the 12 cores process (see figure 1 – right), which must be carried out in a precise order to minimize the discomfort of the patient. The protocol is needed because cancer is generally not visible in the US images; though the samples have to be taken as regularly as possible over the prostate. Additional sites

---

corresponding for example to suspicious zones located during a MRI examination can be added to the standard protocol.

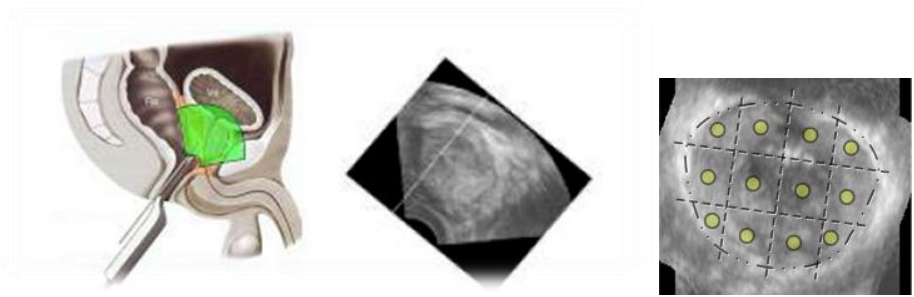

**Figure 1.** US-guided prostate biopsy. (Left) anatomy, (middle) typical ultrasound image, (right) 12-core protocol.

This gesture needs a good 3D representation of the prostate whilst standard US images are 2D. Building an accurate mental 3D representation, well coordinated with 2D US images, requires very good hand-eye coordination which may be difficult to acquire; as a consequence biopsies can quickly be poorly distributed.

The conventional teaching method for this practice is only based on companionship: the student learns by observing an expert doing the operation and then reproduces it. With this method, it may be very difficult to understand the intra-rectal gesture only by observation. The hand-eye coordination and the spatial localization may be hard to acquire and may need a long training, difficult to realize in a clinical environment. Moreover the evaluation of the performance of the trainee is purely qualitative.

The following sections present a new way to teach this gesture using an ultrasound-based prostate biopsy simulator integrated in a complete learning environment with exercises.

## 1. Material and Methods

In order to set up a new way to teach ultrasound guided prostate biopsies, a complete learning environment is being built. The user, most often a urologist resident, must be completely autonomous with the system and should be able to learn everything about prostate biopsy, from theoretical aspects like anatomy of this organ, to practice exercises with simulation of biopsy sequences on specific cases. This learning environment should also provide practical advices about what to do during the practice of this gesture. Such a system aims at proposing a new method to learn this gesture as realistically and completely as possible.

*1.1. The Simulator*

BiopSym simulates ultrasound biopsies of prostate. A first kernel of the system was developed in 2008 [1]. Its principle is to use a force feedback device Sensable Omni

Phantom (cf. Figure 2) to simulate the interaction of the probe with the anatomy of the patients. The stylus of this device represents the ultrasound probe used during this medical gesture. 3D real ultrasound images acquired on real patients are used to compute the visualization. In the new version which graphical user interface has been redesigned, evaluation of the trainee has been included; exercises and educational content have been added.

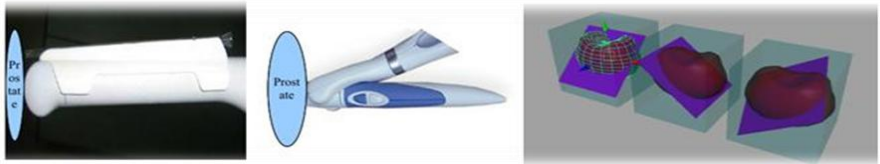

**Figure 2.** (Left) US probe, (middle) Phantom modeling and (right) clipping planes.

The evaluation of the virtual biopsies is made thanks to the prostate segmentation [2]: the issue is to quantify the way the 12 zones of the prostate have been reached and which amount of tissue has been sampled inside the prostate (it is not rare to get samples out of the gland). The segmentation produces a mesh representing the organ and its bounding box; its edges are known. To locate all the sample sites, a simple decomposition of the box into 12 zones is done and comparisons between them and the coordinates of the needles in the simulator validate or not the sample.

*1.2. Architecture*

3D US volumes used in the application were acquired during real biopsy sessions at La Pitiè Salpètrière Hospital, Paris, France. Patient data (age, prostate size, PSA level, etc) are also recorded in a SQL database. BiopSym uses open source libraries to compute the simulation.

Users of this application will be physicians in urology so a significant care was taken to avoid common errors of medical software user interfaces. Lack of consistency and non intuitive design [3],[4],[5] are the most frequent criticisms made to medical software. These properties were considered in order to build user-friendly software with a "look & feel" design (see Figure 3).

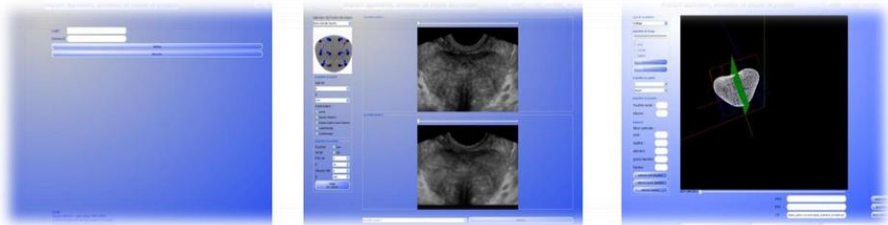

**Figure 3.** Screenshots of BiopSym.

*1.3. Exercises*

The list of exercises was described in a didactic survey based on an analysis of stakes and methods of urologists for prostate biopsies; it was made by a student in educational science of UPMF University working on the project. Its aim is both the acquisition of a good mental representation of the prostate during an echo-guided intervention and the improvement of the user theoretical knowledge. Exercises include: (1) a questionnaire to determine the probability of a cancer regarding patients' data (PSA level, prostate volume, age, etc), (2) image recognition like volume computation on US images (selection of diameters and height) or area localizations (selection of the bladder on a US volume for instance) and (3) several types of simulations with different constraints (volume of the prostate, position of the patient, target, link MRI/US, etc). During simulation, assistance can be provided to help the user; such assistance can consist in adding additional views (coronal, 3D, etc) that the user would not have in a standard biopsy procedure (see Figure 4).

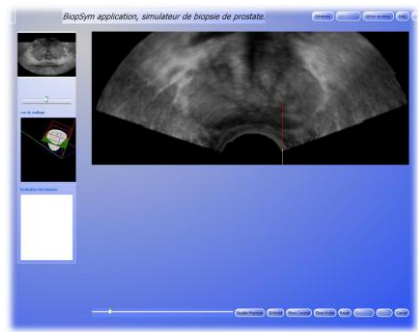

**Figure 4.** Virtual biopsy simulation with assistance.

Evaluations of these exercises allow targeting the weakness of the user and specific exercises to correct them are recommended to him in order to improve his control of this gesture. Thus, the user will be totally autonomous with the application and he will be able to progress alone.

User feedbacks, integrated in the personal panel of the application, display the pedagogical path and evolution of the performances of the user [6],[7]. He is given the possibility to see his timeline representing everything he has done with the application; he can also get details about each activity. For example, he is able to visualize the result of the sequence of biopsies he made two months ago and to watch the results on a 3D view. Charts representing the evolution of his score are also available.

*1.4. Educational Content*

The application also contains a complete lecture about prostate biopsy; it was prepared by a clinical expert. It can be decomposed in two parts:
-  A slideshow where the expert reminds and explains the basic knowledge required to practice prostate biopsy: the anatomy of the prostate and its appearance in ultrasound images, its different components and the realisation of the gesture (see Figure 5).
- Lectures about the practice of biopsy: the way it is realized, the checklist of all the details needed to make it successful and an explanation text from the "Association Française d'Urologie" (French Urologist Association) which explains it for the patient.

The educational part provides a theoretical support to the exercises and helps the user to acquire a solid background about this medical gesture. All the resources can be transferred to a USB pen for the trainee.

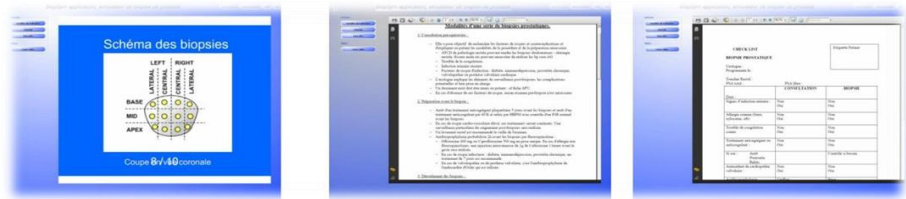

**Figure 5.** Examples of educational content.

## 2. Results

BiopSym allows now the teaching of ultrasound guided prostate biopsy as pertinently and completely as possible. It provides a theoretical support about this gesture and above all a new way to learn this gesture with the simulations. The exercises and their recommendations will help the users to progress and allow them to be totally autonomous. The assistances will help them to get quickly a good mental representation of the prostate and so master faster the biopsy realization. The feedbacks about user performance allow him to visualize his progress and involve him in this practice. BiopSym evaluation with real medical student is scheduled for 2011.